\documentclass{ws-procs975x65} 
\begin{document}
\title{Rare decays at the LHCb experiment}

\author{L. Pescatore$^*$, on behalf of the LHCb collaboration}

\address{$^*$University of Birmingham, Birmingham, UK\\
$^*$E-mail: luca.pescatore@cern.ch}



\begin{abstract}
Rare decays of beauty and charm hadrons offer a rich playground to
make precise tests of the Standard Model and look for New Physics at
the level of quantum corrections. A review of recent LHCb results will be presented.
\end{abstract}


\bodymatter

\section{Introduction and the LHCb detector}

The LHCb experiment\cite{Alves:2008zz} is a forward spectrometer fully
instrumented in the pseudo-rapidity range $2 < \eta < 5$. It is characterised by an 
excellent particle identification, given by two RICH detectors, and good
impact parameter and momenta resolutions combined with
a highly efficient and flexible trigger able to trigger
on muons, electrons, hadrons and photons with low $p_T$ thresholds.
The experiment is very well suited to study rare decays of $b$- and
$c$-hadrons as it benefits from large $b\overline{b}$ and
$c\overline{c}$ cross-sections and can access very low transverse momentum ranges thanks to its forward geometry.

The LHCb detector is a precision machine, designed to test in detail the Standard Model (SM).
In these proceedings the analysis of, so-called, ``electroweak-penguin" (EWP) decays will be described.
These are Flavour Changing Neutral Currents (FCNC), forbidden in the
SM at tree level, but allowed at loop level. 
Therefore they are very sensitive to New Physics (NP) entering
the loops and can probe higher mass scales than direct searches.
Furthermore they offer a rich environment with a wealth of observables sensitive to NP entering in the loops. 
The result of two searches for Lepton Flavour Violating (LFV) decays,
forbidden in the SM but with possible tree level contributions beyond
it, will also be reported.
The analyses presented in these proceedings are based on a dataset
corresponding to up-to 3~fb$^{-1}$ of integrated luminosity:
1~fb$^{-1}$ from 2011 run at a collision energy of 7~TeV and 2~fb$^{-1}$ from 2012 run at 8~TeV.

\section{Branching ratios and angular analysis of $B\rightarrow K^{(*)}\mu\mu$ decays}

The branching fractions (BR) of $B\rightarrow K\mu^+\mu^-$ and $B\rightarrow K^{*}\mu^+\mu^-$ decays are highly sensitive
to NP entering in the loops and LHCb is well suited to study these decays since it can efficiently trigger on muons.
As a first result, the branching fractions of the $B^+\rightarrow K^{*+}\mu^+\mu^-$, $B^+\rightarrow K^{+}\mu^+\mu^-$
and $B^0\rightarrow K_S^{0}\mu^+\mu^-$ decays are determined using an
integrated luminosity of 3~fb$^{-1}$. These are analysed
reconstructing $K^{*+}\rightarrow K^0_S \pi^+$ and $K^0_S \rightarrow
\pi^+\pi^-$ decays.
Results\cite{Aaij:2014pli} are given extrapolating under the $J/\psi$ region which is vetoed in the measurement.
\begin{align}
{\cal B}( B\rightarrow K^+\mu^+\mu^-) &= (4.29 \pm 0.07 \text{(stat)} \pm 0.21 \text{(sys)}) \times 10^{-7}\\
{\cal B}(B\rightarrow K^{*+}\mu^+\mu^-) &= (9.24 \pm 0.93 \text{(stat)} \pm 0.67 \text{(sys)}) \times 10^{-7}\\
{\cal B}(B\rightarrow K_S^{0}\mu^+\mu^-) &= (3.27 \pm 0.34 \text{(stat)} \pm 0.17 \text{(sys)}) \times 10^{-7}
\end{align}

Theoretical predictions for these branching fractions are affected by large errors due to calculations of $B\rightarrow K^{(*)}$ form factors.
Therefore it is important to study quantities where these uncertainties are reduced.
One of these quantities is the isospin asymmetry defined as:
\begin{equation}
A_I = \frac{\mathcal{B}(B^0 \rightarrow K^{(*)0}\mu^+\mu^-) - (\tau_0/\tau_+) \mathcal{B}(B^+ \rightarrow K^{(*)+}\mu^+\mu^-)}
   {\mathcal{B}(B^0 \rightarrow K^{(*)0}\mu^+\mu^-) + (\tau_0/\tau_+) \mathcal{B}(B^+ \rightarrow K^{(*)+}\mu^+\mu^-)},
\label{eq:isospinAsym}
\end{equation}
where $\tau_0$ and $\tau_+$ are the lifetimes of $B^0$ and $B^+$.
\begin{figure}
\label{fig:isoAsym}
\centering
\includegraphics[width=0.49\textwidth]{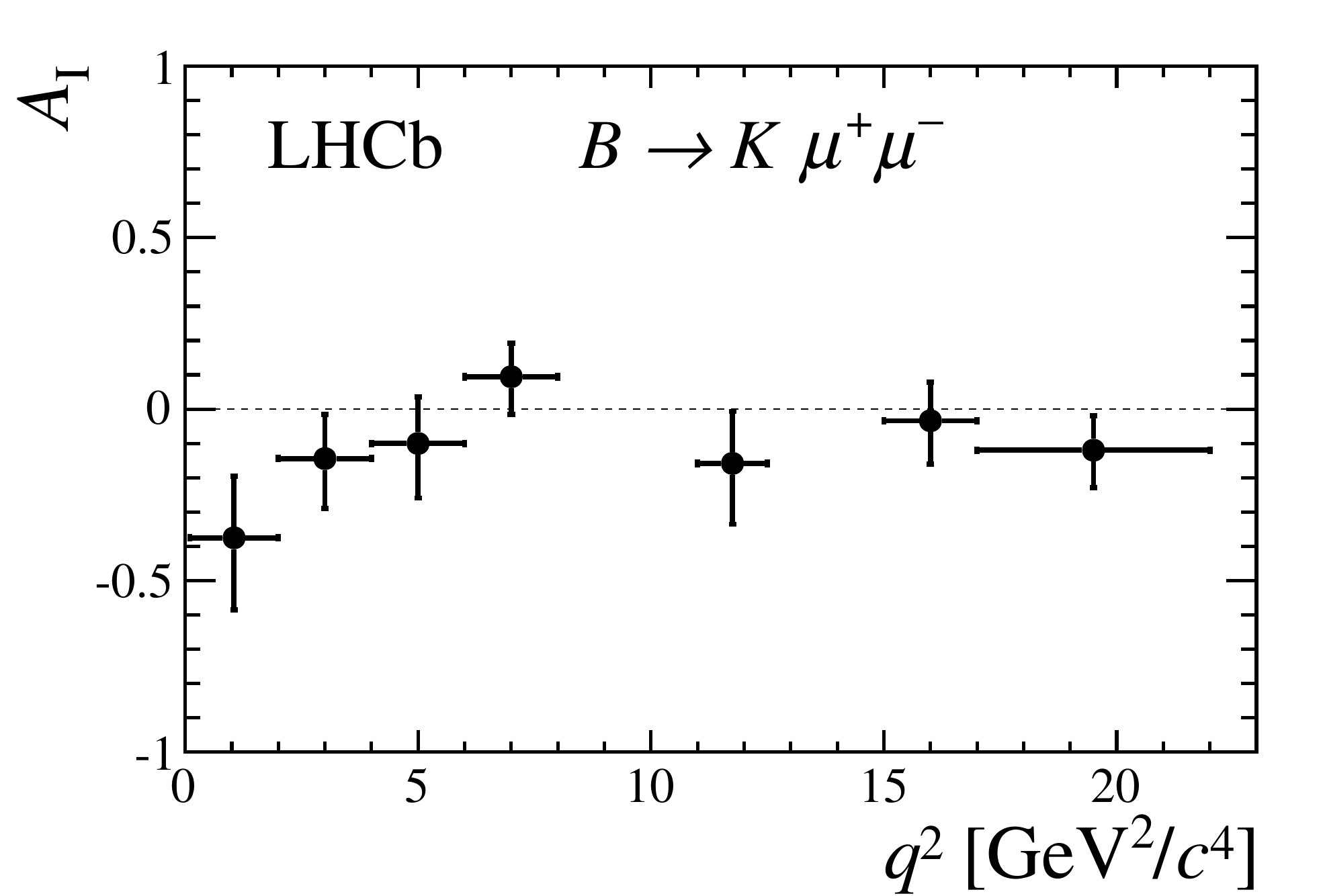}
\includegraphics[width=0.49\textwidth]{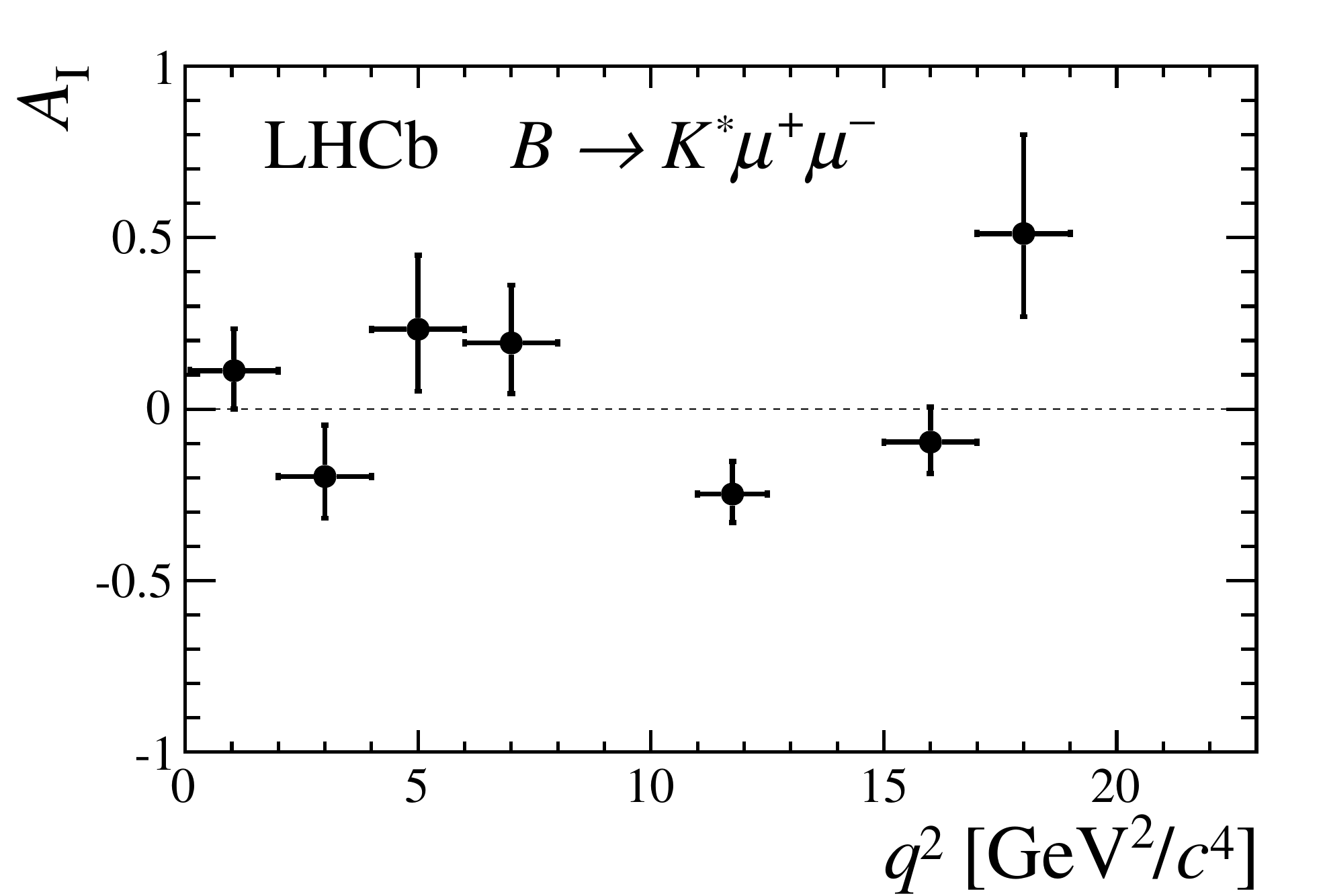}
\caption{Isospin asymmetry in bins of $q^2$ for the $K$ channels
  (left) and $K^*$ channels (right).}
\label{fig:isoAsym}
\end{figure}

This corresponds to analysing the differences in $b\rightarrow s$
transitions where the spectator quark is a $u$- or $d$- type quark.
The value for the isospin asymmetry is expected to be $O(1\%)$ in the SM, due to correction of the order $\sim m_q/m_b$.
 
One effect which could bias the result is the production asymmetry of charged and neutral $B$ mesons.
For this reason the BR are normalised using the resonant channels where the dimuon final
states come from $J/\psi$ decays.
In this way the production asymmetry cancels but on the other hand null asymmetry between $J/\psi$ channels is assumed.
 
Figure \ref{fig:isoAsym} shows isospin asymmetries for the $K$ and
$K^*$ cases in bins of the dimuon invariant mass, $q^2$.
The null hypothesis is tested against the simplest alternative, a constant value that differs from zero.
The result is found to be in agreement with the SM within $1.5\sigma$\cite{Aaij:2014pli}.

\begin{figure}
\label{fig:AngAna}
\centering
\includegraphics[width=0.52\textwidth]{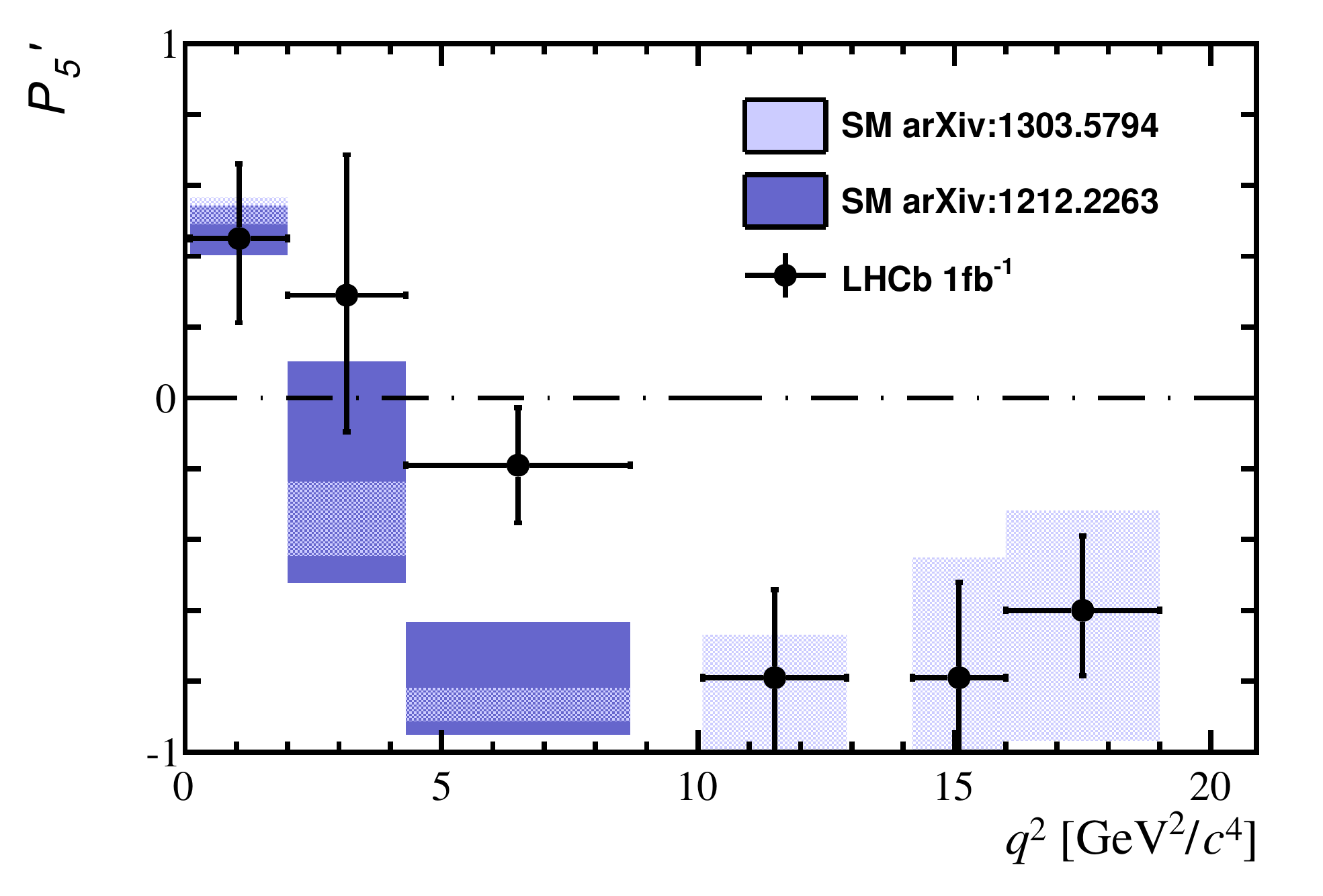}
\caption{$P'_{5}$ variable as a function of $q^2$ with theoretical predictions. }
\end{figure}

Finally, an angular analysis is performed on these decays. Several angular observables have been proposed in the literature,
 that have a reduced dependence on form factors. LHCb has measured four observables, named $P'_{i=4,5,6,8}$ in $q^2$ bins.
While most of the quantities are found to be in good agreement with
the SM, a local $3.7\sigma$ discrepancy with respect to SM
predictions is observed in one $q^2$ bin for the $P_5'$ observable. The experimental measurement of $P_5'$\cite{Aaij:2013qta} along with
the SM prediction\cite{Descotes-Genon:2013vna} are shown in Fig.~\ref{fig:AngAna}.

\section{Testing lepton universality}

Lepton universality is the equality of electroweak couplings for all leptons.
The analysis here presented tests lepton universality between electrons and muons
using penguin decays, where NP in the loops could break the symmetry. This is done by
measuring the ratio
\begin{equation}
R_K = \frac{\int^{q^2_{max}}_{q^2_{min}} \frac{ d\Gamma(B^+\rightarrow K^+ ee)}{dq^2}dq^2 }
        {\int^{q^2_{max}}_{q^2_{min}} \frac{ d\Gamma(B^+\rightarrow
            K^+ \mu\mu)}{dq^2}dq^2 }.
\end{equation}

The analysis is done in the $1-6$~GeV$^2$ region in $q^2$, where theory
calculations are expected to be most reliable.
The quantity measured is actually the double ratio with respect to the 
corresponding charmonium decays.
These channels have same final daughters and similar kinematics, reducing systematic uncertainties.

The challenge of this analysis are the electronic channels, since
electron reconstruction is more challenging due to material
interactions and multiple scattering.
To maximise the yield three trigger categories are considered: events triggered by the electron, by the kaon
and by other particles in the event. Furthermore, electrons can radiate bremsstrahlung photons, which can affect the
momentum measurement. Therefore an algorithm was developed to recover this energy by looking at calorimeter hits
around the electrons.
Assuming universality holds, the ratio should be close to 1 in the SM with corrections $O(10^-3)$ due to Higgs contributions.
On 3~fb$^{-1}$ of integrated luminosity the ratio is measured to be $0.745^{+0.090}_{-0.074} (\text{stat})
{}^{+0.036}_{-0.036} (\text{sys})$ corresponding to a $2.6\sigma$ deviations from the SM\cite{Aaij:2014ora}. 

\section{Photon polarisation in $B^\pm\rightarrow
  K^\pm\pi^\pm\pi^\mp\gamma$ on 3~fb$^{-1}$}
\label{sec:pipigamma}
\begin{figure}
\centering
\includegraphics[width=0.46\textwidth]{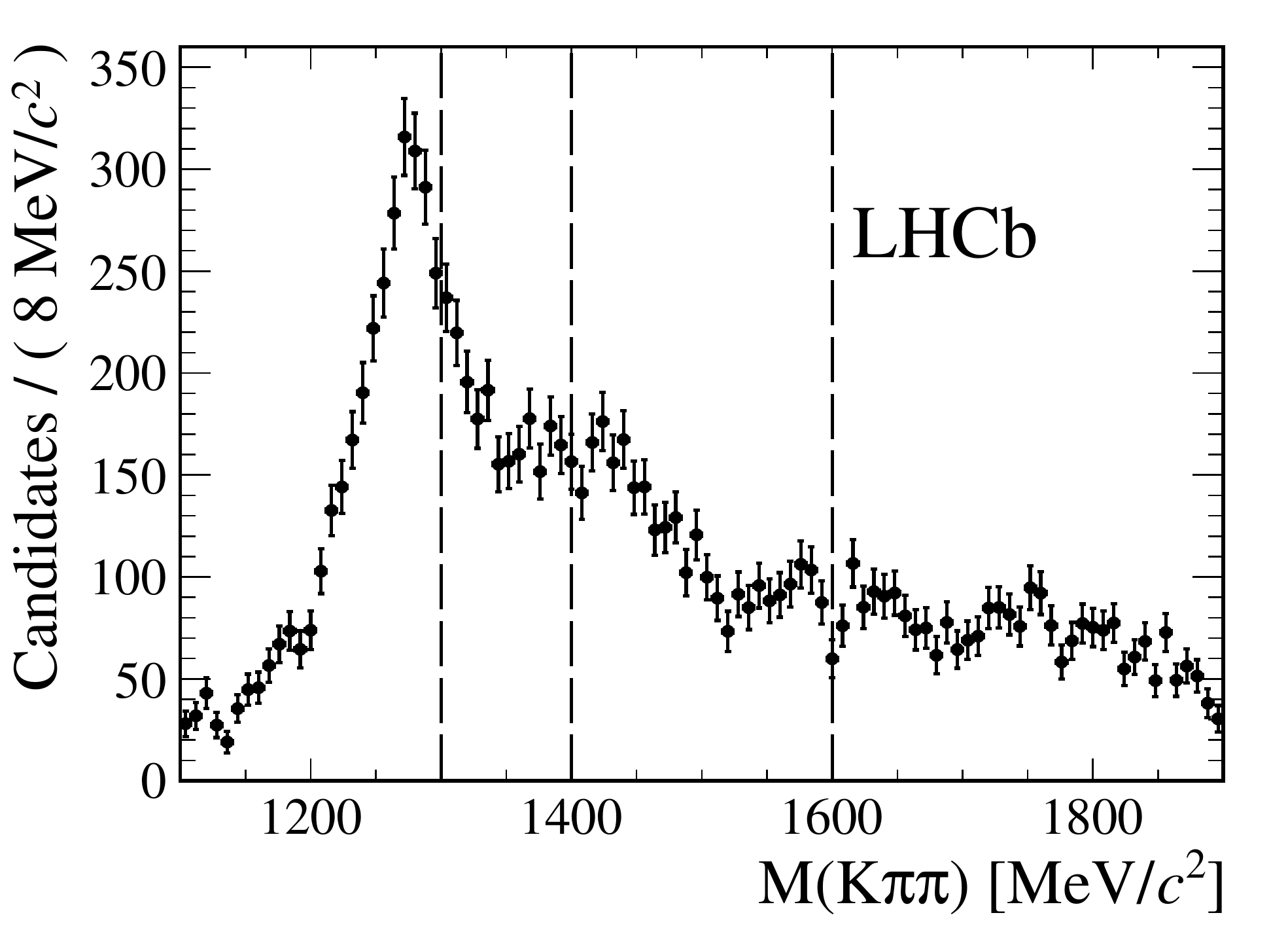}
\includegraphics[width=0.52\textwidth]{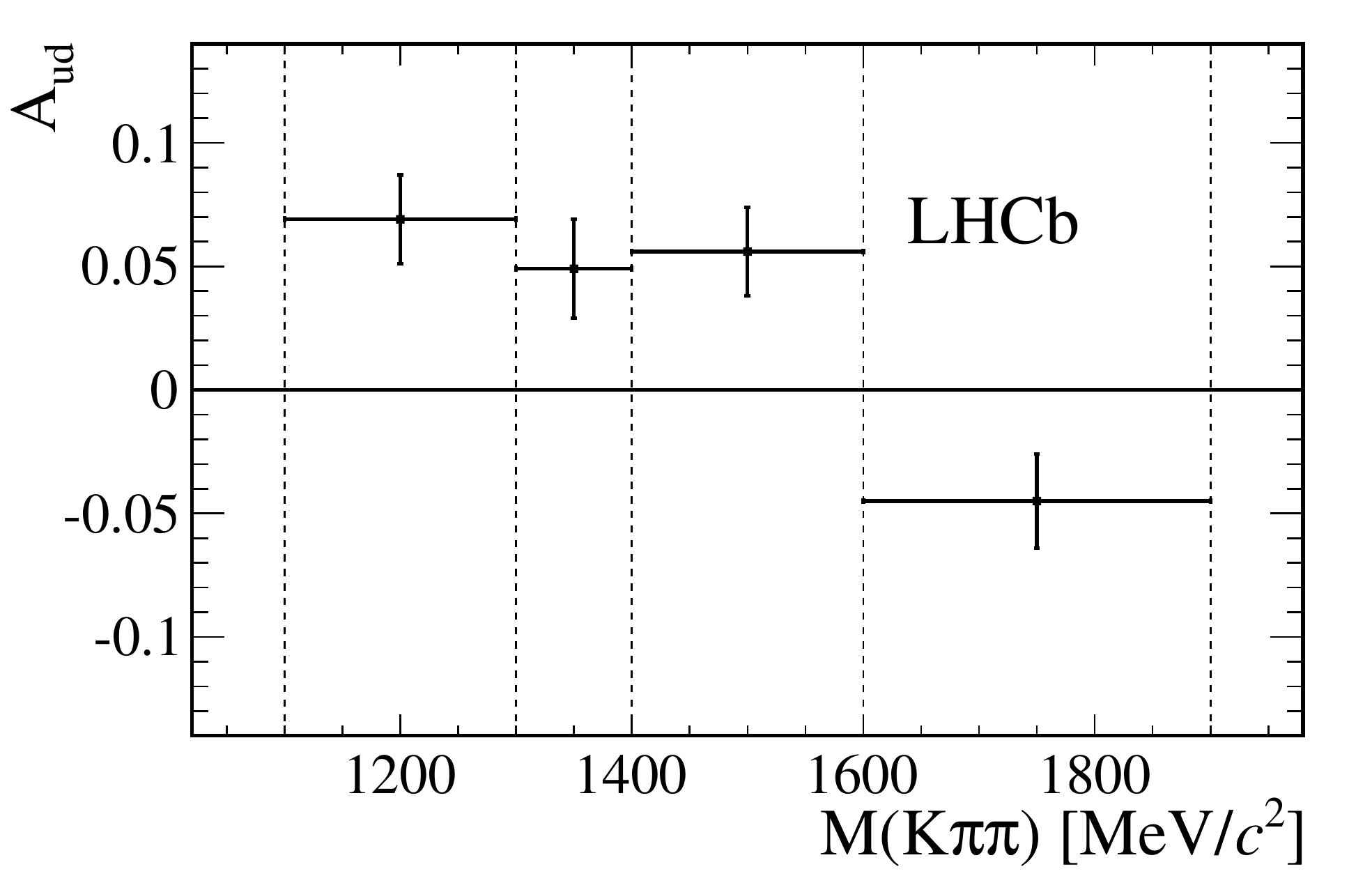}
\caption{$K\pi\pi$ invariant mass (left) and up-down asymmetry in bins
of $K\pi\pi$ mass (right).}
\label{fig:Aupasym}
\end{figure}

The $B^\pm\rightarrow K^\pm\pi^\pm\pi^\mp\gamma$ decay is a $b\rightarrow s$ transition mediated by an EWP loop. The SM predicts that the photon
emitted from the loop is predominantly left-handed, since the recoiling $s$ quark, that couples to a W boson, is left-handed. In several 
extensions of the SM the photon acquires a significant right-handed component due to the exchange of heavy fermions.
The measurement is performed by studying the distribution of the angle
$\cos\theta$, between the photon and the plane formed by the three hadrons.
In a second step the up-down asymmetry, proportional to the photon polarisation, is extracted from the distribution. 
One of the challenges of the analysis is the interpretation of the
result due to the limited knowledge of the composition of the
$K\pi\pi$ final state, where different resonances contribute. To maximise the information given the measurement is
done in 4 bins of $K\pi\pi$ invariant mass,
$m(K\pi\pi)$. Figure \ref{fig:Aupasym} shows the $m(K\pi\pi)$
distribution and the up-down asymmetry in bins of $m(K\pi\pi)$.
Combining the 4 bins results in the first observation of a parity
violating polarisation in a $b\rightarrow s\gamma$ transition\cite{Aaij:2014wgo} with a
significance of $5.2\sigma$.
A deeper understanding of the $K\pi\pi$ system is needed to understand
if this asymmetry is consistent with the pre-dominantly left-handed
polarisation expected in the SM.

\section{$\tau^-\rightarrow\mu^-\mu^+\mu^-$ on 1~fb$^{-1}$}

Lepton flavour conservation is well established in the SM by measuring
decays such $\mu\rightarrow e \gamma$ and $\mu\rightarrow eee$.
However it has no strong theoretical justifications. In fact we already know that oscillations of neutrinos in loops can
generate LFV. However, since neutrinos oscillate over long distances, this has a very low BR \mbox{($\sim10^{-40}$)}.
But New Physics 
can increase it up to our reach, O($10^{-9}$).
The number of signal events observed is compatible with the number
of background expected. Therefore an upper limit\cite{Khanji:2014dca}
is set using the CLs method: 
\mbox{$\mathcal{B}(\tau\rightarrow\mu\mu\mu) < 8.0(9.8) \times 10^{-8} \text{ at } 95(90)\% \text{ CL}$}.
This is the first limit on this decay at a hadron collider and competes with limits set at the B-factories.

\section{Search for Majorana neutrino in $B^-\rightarrow \pi^+\mu^-\mu^-$ on 3~fb$^{-1}$}

Neutrinos can be Majorana (if $\nu = \overline{\nu}$) or Dirac fermions.
Searches for Majorana neutrinos looking at neutrino-less
double-$\beta$ decays have found no signal yet.
Therefore it is interesting to look for them in B decays using a different approach.
The $B^-\rightarrow \pi^+\mu^-\mu^-$ decay is forbidden in the SM
because it violates lepton number conservation but it can happen though
the exchange of a Majorana neutrino.
The analysis is done using two different selections optimised for neutrino
lifetimes ($\tau_N$) shorter and longer than 1~ps. In fact for
lifetimes above 1~ps the neutrino
has time to travel in the detector and one can use the extra information coming from the displaced vertex.
The number of events found is compatible with the number of events
expected in the background only hypothesis. Therefore an upper limit
is set\cite{Aaij:2014aba}: $\mathcal{B}(B^-\rightarrow \pi^+\mu^-\mu^-) < 4.0 \times 10^{-9} \text{ at } 95\% \text{ CL}$,
for $\tau_N < 1$~ps. In the article also model dependent upper limits
are set for lifetimes above 1~ps.

\section{Conclusions}
Rare $B$ and $D$ decays proceeding through FCNC are sensitive probes for New Physics effects.
These decays are extensively studied at LHCb and allow sensitive tests of the SM.
Most results are found to be compatible with the SM with interesting
deviations in the $R_K$
measurement and the $P'_5$ variable in the
$B^0\rightarrow K^{*0}\mu^+\mu^-$ angular analysis. 
Worth mentioning is also the first measurement of a parity-violating photon polarisation in
$b\rightarrow s\gamma$ transitions with $5.2\sigma$ significance.
Finally, upper limits competing with the B-factories are set on LFV decays, where no signal is found.

\bibliographystyle{ws-procs975x65}
\bibliography{references}

\begin{thebibliography}{1}

\bibitem{Alves:2008zz}
J.~Alves, A.~Augusto {\em et~al.}, {The LHCb Detector at the LHC}, {\em JINST}
  {\bf 3}, p. S08005  (2008).

\bibitem{Aaij:2014pli}
R.~Aaij {\em et~al.}, {Differential branching fractions and isospin asymmetries
  of $B \to K^{(*)} \mu^+ \mu^-$ decays}, {\em JHEP} {\bf 1406}, p. 133
  (2014).

\bibitem{Aaij:2013qta}
R.~Aaij {\em et~al.}, {Measurement of Form-Factor-Independent Observables in
  the Decay $B^{0} \to K^{*0} \mu^+ \mu^-$}, {\em Phys.Rev.Lett.} {\bf 111}, p.
  191801  (2013).

\bibitem{Descotes-Genon:2013vna}
S.~Descotes-Genon, T.~Hurth, J.~Matias and J.~Virto, {Optimizing the basis of
  ${B} \to {K}^{*}\ell^+ \ell^-$ observables in the full kinematic range}, {\em
  JHEP} {\bf 1305}, p. 137  (2013).

\bibitem{Aaij:2014ora}
R.~Aaij {\em et~al.}, {Test of lepton universality using $B^{+}\rightarrow
  K^{+}\ell^{+}\ell^{-}$ decays}  (2014).

\bibitem{Aaij:2014wgo}
R.~Aaij {\em et~al.}, {Observation of photon polarization in the $b \to
  s\gamma$ transition}, {\em Phys.Rev.Lett.} {\bf 112}, p. 161801  (2014).

\bibitem{Khanji:2014dca}
B.~Khanji, {Searches for lepton flavour violation and lepton number violation
  at LHCb}, {\em Nucl.Phys.Proc.Suppl.} {\bf 248-250}, 91  (2014).

\bibitem{Aaij:2014aba}
R.~Aaij {\em et~al.}, {Search for Majorana neutrinos in $B^- \to
  \pi^+\mu^-\mu^-$ decays}, {\em Phys.Rev.Lett.} {\bf 112}, p. 131802  (2014).

\end{thebibliography}

\end{document}